\documentclass{INTERSPEECH2023}
\usepackage[flushleft]{threeparttable}
\usepackage{multirow}
\usepackage{makecell}
\usepackage{pifont}
\usepackage{amsmath}
\usepackage{booktabs} 
\usepackage{algorithm}
\usepackage{algpseudocode}
\usepackage{graphicx}
\usepackage{enumitem}

\newcommand{\cmark}{\ding{51}}
\newcommand{\xmark}{\ding{55}}

\title{MixRep: Hidden Representation Mixup for Low-Resource Speech Recognition}
\name{Jiamin Xie, John H.L. Hansen}
\address{
    Center for Robust Speech Systems (CRSS), University of Texas at Dallas, TX, 75080}
\email{\{Jiamin.Xie, John.Hansen\}@utdallas.edu}

\begin{document}

\maketitle
 
\begin{abstract}
In this paper, we present MixRep, a simple and effective data augmentation strategy based on mixup for low-resource ASR. MixRep interpolates the feature dimensions of hidden representations in the neural network that can be applied to both the acoustic feature input and the output of each layer, which generalizes the previous MixSpeech method. Further, we propose to combine the mixup with a regularization along the time axis of the input, which is shown as complementary. We apply MixRep to a Conformer encoder of an E2E LAS architecture trained with a joint CTC loss. We experiment on the WSJ dataset and subsets of the SWB dataset, covering reading and telephony conversational speech. Experimental results show that MixRep consistently outperforms other regularization methods for low-resource ASR. Compared to a strong SpecAugment baseline, MixRep achieves a +6.5\% and a +6.7\% relative WER reduction on the eval92 set and the Callhome part of the eval'2000 set.

\end{abstract}
\section{Introduction}
Deep learning research has fueled many recent advancements toward solving the automatic speech recognition (ASR) task. The end-to-end (E2E) ASR \cite{chanLAS,ctc,rnnt} predicts the textual output from the time-frequency input by a deep stack of convolutional neural networks (CNN) \cite{cnn}, recurrent neural networks (RNN) \cite{rnn}, or attention layers \cite{att}. The large modeling capacity of the E2E ASR model helps learn a direct mapping from the input to the output sequence effectively, as shown in many works \cite{chan2020imputer,multilingual}. While large models are powerful to achieve impressive performance \cite{tuske21_interspeech} given a sizeable training set, they tend to memorize examples and become overly confident with incorrect predictions \cite{zhang2018mixup, verma2019manifold}. For low-resource scenarios, overfitting becomes an issue \cite{sharma20c_interspeech,peterson21_interspeech} with other challenges like diverse acoustic variations \cite{apollo,Chen2021ScenarioAS} and language mismatch \cite{comprLR, size_fit}.


Data augmentation is one effective way to expand the training data and make models generalize \cite{wu2020generalization, w2v_robust}. Developed techniques for ASR create multiple views of the original speech \cite{danperturb} by applying vocal tract length normalization \cite{jaitly2013vocal}, reverberation \cite{reverb}, and tempo variations \cite{danperturb}. Advanced methods synthesize speech directly using the state-of-the-art text-to-speech \cite{tts_model} and voice conversion \cite{cyclegan_vc} models, which is shown beneficial for low-resource distant talks \cite{tsunoo2021data}. Other methods like SpecAugment \cite{specaug} randomly crops and modifies the input spectrogram like images along both time and frequency dimensions. Feature mixup \cite{zhang2018mixup,verma2019manifold} is another angle to create artificial examples by exploring the input space through interpolation, where a mixup refers to the convex combination of two training features. One recent work of ASR studies the mixup between mel-spectrograms of two utterances and trains the E2E model to predict both reference texts from the mixed feature \cite{meng2021mixspeech}.

Since the hidden representation space of an ASR model can encode information (e.g. phoneme, word, and 
semantics) more abstract than the acoustic features at the input \cite{belinkov2017analyzing,li2020does}, we reason performing the mixup of hidden representations is beneficial. As shown in the previous study \cite{verma2019manifold}, the mixup performed at deep layers of a model has regularization effects on the representations. It reduces variations in the dimensions that encode redundant information and also smooths the classification boundaries among representations, which alleviates over-confident predictions for adversarial or ambiguous input. For E2E speech recognition, we hypothesize such regularization would improve the overall learning as the speech input contains many variations caused by low-dimensional factors such as content, speakers, and channels \cite{wang21n_interspeech}.

In this study, we propose a data augmentation method for low-resource ASR based on representation mixup, named MixRep. The contribution of this work is as follows,
\begin{enumerate}
    \item A data augmentation strategy using the mixup of hidden representations for low-resource speech recognition \footnote{\url{https://github.com/jiamin1013/mixrep-espnet}}
    \item Highlight of the complementary regularization on both time and frequency (feature) dimensions for mixup methods
    \item Investigation of other techniques, e.g. SpecAugment \cite{specaug} and MixSpeech \cite{meng2021mixspeech}, and their comparison to MixRep
\end{enumerate}



\section{Related Work}
The concept of input mixup \cite{zhang2018mixup} has been successfully applied to classification tasks because the labels are one-hot and easy for interpolation, e.g. pictures \cite{wang2020neural}, acoustic scenes \cite{pham2019robust}, speakers \cite{mixup_spkverify}, etc. For ASR acoustic model training \cite{medennikov2018investigation}, the mixup is conducted for the HMM state labels aligned to the speech input. For tasks with label sequences of different lengths, the mixup of training losses is used instead, e.g. for the E2E model training in speech recognition \cite{meng2021mixspeech} or machine translation \cite{guo2020sequence}. The Manifold Mixup \cite{verma2019manifold} extends input mixup to the hidden representations of a deep neural network, which is the focus of our study. For speech input, this has only been previously studied for sound classification \cite{jindal20_interspeech} and one recent work on speech translation \cite{fang2022stemm}, where the latter applies mixup to representations from two modalities and does not consider a mixup of target sequences. Unlike previous work, we investigate the application of Manifold Mixup to train an E2E ASR model. We intend to learn the behavior of different layers, so we do not search layer combinations as extensively as done in \cite{jindal20_interspeech}. Our approach is similar to the MixSpeech \cite{meng2021mixspeech} method but extends it and explores the combination of techniques. 

\section{Method}
In this section, we first review the mixup \cite{verma2019manifold} concept.
We then explain the MixSpeech method \cite{meng2021mixspeech} that applies mixup to E2E ASR. Finally, we describe our proposed method which extends the speech mixup to the hidden representation, and mention its regularization effect on the feature dimension.
\subsection{Manifold Mixup}
The Manifold Mixup \cite{verma2019manifold} is a generalized version of the input mixup \cite{zhang2018mixup} that allows representation ouput from any layer of a neural network model to be linearly interpolated (i.e. mixup). For an arbitrary $K$-layers model, we denote $f_{n,k}(\cdot)$ the underlying function that processes data from the $n$-th layer input to the $k$-th layer output, where $n=0$ is the model input and $f_{0,0}(\cdot)$ is the identity function. Suppose a supervised learning task has input features $X$ and one-hot labels $Y$, the Manifold Mixup trains the model by mixing up the hidden representations and labels,
\begin{gather}
    R_{k} = \lambda*f_{0,k}(X_{i}) + (1-\lambda)*f_{0,k}(X_{j}), \label{eqn:1} \\
    Y_{mix} = \lambda*Y_{i} + (1-\lambda)*Y_{j}, \\
    \mathcal{L}_{mix} = \mathcal{L}(f_{k,K}(R_{k}), Y_{mix}),
\end{gather}
where $\lambda \in [0,1] \sim Beta(\alpha, \alpha) $ with $\alpha \in (0, \infty)$ and $i$ and $j$ denote two training examples. The interpolation results in a new training example represented by the hidden dimensions of the model, thus it is an effective data augmentation method. We note the input mixup \cite{zhang2018mixup} becomes a special case of the Manifold Mixup \cite{verma2019manifold} when $n$ and $k$ are both 0. 

\subsection{MixSpeech: Input Mixup}
MixSpeech \cite{meng2021mixspeech} is a data augmentation method developed for E2E ASR training based on the input mixup \cite{zhang2018mixup}. For a pair of utterances, this method mixes up acoustic features of these utterances in the frequency dimensions frame-by-frame. Because speech input and text output have different lengths with the alignment unknown, mixing two word labels at the same position does not correspond to a simultaneous time when both words are spoken. So, the MixSpeech interpolates the losses of recognizing each textual label sequence instead. 


\subsection{MixRep: Hidden Representation Mixup}
\begin{algorithm}[h]
    \caption{Hidden Representation Mixup (MixRep)}
    \label{algo1}
    \begin{algorithmic}[1]
        \State \textbf{Given} a subset $\mathcal{S} \in \{0,1,\ldots,K\}$, a beta coefficient $\alpha$, a pre-processing function $m(\cdot)$
        \Procedure{mixup}{$x,y,\lambda$}
        \State get $batchSize$ from $x$
        \State $indArr$ $\leftarrow$ shuffle list [$0,1,\ldots,batchSize-1$]
        \State $x \leftarrow \lambda*x + (1-\lambda)*x[indArr,:]$ \hfill // \textit{interpolation}
        \State $\Tilde{y} \leftarrow\ y[indArr,:]$
        \State \Return $x$, $\Tilde{y}$
        \EndProcedure
        \For {$each\ batch$} 
        \State $\lambda \sim Beta(\alpha, \alpha)$ \hfill // \textit{sample an interpolation weight}
        \State $k \sim Uniform(\mathcal{S})$ \hfill //  \textit{sample a layer index}
        \State $x \leftarrow batch$
            \For {$(index,\ layer)$ in $layers$}
                \If{$index = k$}
                \State $x,\ \Tilde{y} \leftarrow$ \Call{mixup}{$x,y,\lambda$}
                \EndIf
                \If{$index = 0$}
                \State $x \leftarrow m(x)$ \hfill// \textit{for masked-based preprocessing}
                \EndIf
                \State $x \leftarrow layer.forward(x)$
            \EndFor
        \State backward $loss \leftarrow \lambda*\mathcal{L}(x, y) + (1-\lambda)*\mathcal{L}(x,\Tilde{y})$ 
        \EndFor
    \end{algorithmic}
\end{algorithm}
We propose MixRep to create artificial examples during training by mixing hidden representations of an E2E ASR model, inspired by the previous methods \cite{verma2019manifold, meng2021mixspeech}. Reusing $R_{k}$ defined in Equation \ref{eqn:1}, MixRep interpolates sampled utterances $i$ and $j$ frame-by-frame by their respective output from the $k$-th layer of a model. For the textual label sequences $Y$, MixRep trains the model to optimize the following loss,
\begin{equation}
\begin{split}
    \mathcal{L}_{mixRep} =\ &\lambda *\mathcal{L} (f_{k,K}(R_{k}), Y_{i}) \\ & + (1-\lambda)*\mathcal{L}(f_{k,K}(R_{k}), Y_{j}),
    \label{eqn:2}
\end{split}
\end{equation}
where $k$ is drawn uniformly from a set of eligible layers $S$ on each forward pass. When $k=0$, since the hidden representations are mel-spectrograms from the input, MixRep naturally extends the MixSpeech \cite{meng2021mixspeech} method. We present the detailed steps of our proposed method in Algorithm \ref{algo1}.

One key aspect of the mixup methods \cite{zhang2018mixup, verma2019manifold} is their regularization benefits on the feature dimension, aside from data augmentation. By making the interpolation weight in the mixup of features and that of the reference labels match, the method constructs a linear association between the input and output space of the neural network \cite{zhang2018mixup}. For Manifold Mixup \cite{verma2019manifold}, the linearity is constructed for the hidden representation space. This has shown to regularize the feature dimensions of the hidden representations by capturing salient low-dimensional variations and enforcing smooth classification boundaries for predictions made on the representations. Because MixRep regularizes the representation space but speech contains both time and frequency information, we propose the following two configurations of the MixRep method:
\begin{itemize}[leftmargin=1cm]
    \item \textit{Basic}: does not apply any regularization along the time axis of the input, similar to \cite{meng2021mixspeech}
    \item \textit{Time enhanced}: applies regularization along the time axis of the input (e.g. time masking or warping, etc.).
\end{itemize}
To explore the \textit{Time enhanced} approach, we investigate applying regularization to the input (line 18 of Algorithm 1). For deep layers of the model (a large \textit{k}), the representation encodes much information due to a large receptive field. Masking representations at a deep layer then impacts performance since the masked content can be hardly recovered by the limited modeling capacity which follows. In order to recognize the missing content from masking, applying time regularization to the input is effective for helping the following attention-based layers to learn strong representation that captures meaning than fine details from the input. We consider it is crucial for MixRep since a good hidden representation space needs to be established. 

\section{Experimental Setup}
To examine the effectiveness of MixRep, we conduct experiments on ASR benchmarks that evaluate speech from reading newspapers or conversations over the telephone. For the Conformer architecture illustrated in section 4.2, we mix representations from the output of an encoder layer (i.e. after the final LayerNorm \cite{gulati_conformer}) and use the original positional encoding without mixup. We establish SpecAugment \cite{specaug} as our baselines, which randomly and partially masks out time and frequency content from the input. By mixing the input acoustic feature, we recreate the MixSpeech \cite{meng2021mixspeech} method. For fair comparisons, we test both the \textit{Basic} and \textit{Time enhanced} configurations of these methods in our experiments. We then apply the best configuration to mix representations and compare the performance of MixRep to the SpecAugment baseline and the effective MixSpeech.

\subsection{Datasets}
The Wall Street Journal (WSJ)  \cite{paul1992design} and Switchboard (SWB) \cite{godfrey1992switchboard} datasets are investigated in our study. The WSJ dataset includes read speech with transcripts drawn from the newspaper. The data is partitioned into 81 hours of training speech (\textit{si284}), 1 hour for development (\textit{dev93}), and 0.7 hour for evaluation (\textit{eval92}). The SWB dataset contains spontaneous speech from two sides of a conversation over the telephone line. To simulate a low-resource setup, we randomly sample the training data into two subsets totaling 40 hours and 80 hours. We use the single-fold train split without any speed or noise perturbation. We use the eval'2000 (LDC2002S09) dataset as evaluation for SWB, where there are Switchboard (swb) and Callhome (chm) parts that are unseen from the SWB training/validation set.

\subsection{E2E ASR model}
For ASR experiments, we follow recipes provided in the ESPnet toolkit \cite{watanabe2018espnet} to train an E2E ASR model for each dataset, which is further referred to as the \textit{Default} setup. Our models use the listen, attend, and spell (LAS) architecture \cite{chanLAS} that include the Conformer encoder \cite{gulati_conformer} and the Transformer \cite{vaswani2017attention} decoder. We extract 80 mel-filterbanks and 3-dimensional pitch features. The input is then passed through an optional SpecAugment \cite{specaug}, followed by 2D-CNNs with a downsampling factor of 4. The SpecAugment uses time warping with a window size of 5,  two frequency masks with $F=30$, and two time masks with $T=40$, unless otherwise stated. The encoder has 12 layers. The decoder has 6 layers and connects to a softmax layer followed by the cross-entropy (CE) loss. The model is trained jointly by $L_{joint} = \alpha*L_{ctc}+(1-\alpha)*L_{ce}$ \cite{kim2017joint}, where $\alpha$ is set to $0.3$ in our study. The label smoothing weight is $0.1$. The model dimension is 256. The attention modules have 4 attention heads and 2048 linear units with a dropout $p=0.1$. We use 
the warmup learning rate scheduler for all datasets. The learning rate of WSJ peaks at $0.005$ after $30$k steps and that of SWB peaks at $0.006$ after $25$k steps. We use character as output to train the WSJ model and byte-pair-encoding (bpe) with 2000 subword units \footnote{The bpe model is obtained from texts in full SWB training} for the SWB model. The number of elements in a batch is $2.5$M for WSJ and $10$M for SWB. The gradients accumulation is 6 times. We use a CNN kernel size of 15 for WSJ and 31 for SWB. The WSJ is trained for 150 epochs and 300 epochs for SWB. Both experiments finish in 1 day using two or four 2080Ti GPUs.

\subsection{Parameters of MixRep}
We use the beta distribution with a coefficient $\alpha=2$ for all experiments using MixRep. This corresponds to a convex-shaped probability distribution with mean equals 0.5 (i.e. $E[\lambda]=0.5$) and about half of the probability mass (56\%) falls between 0.3 and 0.7. Following MixSpeech \cite{meng2021mixspeech}, we also use $\tau=0.15$ for WSJ (means 15\% data of a batch uses the mixup), but we find $\tau=0.45$ to be more suitable for SWB. Since searching all subsets of the layers in the ASR encoder is infeasible (i.e. $2^{12}=4096$ combinations), we employ the following heuristic: we first apply MixRep to every single layer of the ASR encoder and gather its performance; we then test the set $S$ containing the best-performing layer and the input layer. We report every single-layer performance in section 5.4.

%



\section{Results}
\subsection{Baselines and Previous Methods}
Because the ESPnet default setting includes the SpecAugment, we expect it to be the best and make it the baseline. To make a fair comparison to the \textit{Time enhanced} configuration, we investigate turning off frequency masking for SpecAugment. The original MixSpeech is applied to the Transformer model, so we recreate their method for the Conformer model. The results of these systems are illustrated in Table 1.
\begin{table}[h]
  \caption{WER of baselines and previous methods. T and F refer to SpecAugment regularization along the time and frequency fimension, respectively. Default setup is explained in Section 4.}
  \label{tab:baselines}
  \centering
  \scalebox{0.77}{
  \begin{tabular}{ c l c c c c | c c }
    \toprule
    \multirow{2}{*}{\centering Dataset} &
    \multirow{2}{*}{\centering Model} & \multirow{2}{*}{\centering T} &
    \multirow{2}{*}{\centering F}
    & \multicolumn{2}{c}{\textbf{With LM (\%)}} & \multicolumn{2}{c}{\textbf{No LM (\%)}}\\
    & & & & dev & eval & dev & eval \\
   \midrule
    \multirow{9}{*}{\centering WSJ}& \textbf{Transformer} & & & & & \\
    & Espnet \cite{watanabe2018espnet} & \cmark & \cmark & 7.4 & 4.9 & - & - \\
    & MixSpeech \cite{meng2021mixspeech} & \xmark & \xmark & - & 4.7 & - & - \\
    \cmidrule{2-8}
    & \textbf{Conformer} & & & & & \\
    & \multirow{1}{*}{Default} & \cmark & \cmark & 7.1 & 4.7 & 11.2 & 8.9 \\
    & \multirow{1}{*}{Default} & \cmark & \xmark & \textbf{6.2} & 4.3 & 10.4 & 7.7 \\
    & \multirow{1}{*}{+ MixSpeech (Ours)} & \xmark & \xmark & 6.8 & 4.5 & 10.7 & 8.4 \\
    & \multirow{1}{*}{+ MixSpeech (Ours)} & \cmark & \xmark & 6.3 & \textbf{4.2} & \textbf{9.8} & \textbf{7.5} \\
    \midrule
    \multirow{5}{*}{\centering SWB 40hr} & \textbf{Conformer} & & & & & \\
    & \multirow{1}{*}{Default} & \cmark & \cmark & - & - & 21.3 & 34.1 \\
    & \multirow{1}{*}{Default} & \cmark & \xmark & - & - & \textbf{18.5} & 31.6 \\
    & \multirow{1}{*}{+ MixSpeech (Ours)} & \xmark & \xmark & - & - & 20.7 & 33.0 \\
    & \multirow{1}{*}{+ MixSpeech (Ours)} & \cmark & \xmark & - & - & 18.9 & \textbf{30.6} \\
    \midrule
    \multirow{5}{*}{\centering SWB 80hr} & \textbf{Conformer} & & & & & \\
    & \multirow{1}{*}{Default} & \cmark & \cmark & - & - & 13.5 & 23.3 \\
    & \multirow{1}{*}{Default} & \cmark & \xmark & - & - & 13.2 & 23.3 \\
    & \multirow{1}{*}{+ MixSpeech (Ours)} & \xmark & \xmark & - & - & 14.7 & 25.2 \\
    & \multirow{1}{*}{+ MixSpeech (Ours)} & \cmark & \xmark & - & - & \textbf{13.0} & \textbf{22.6} \\
    \bottomrule
  \end{tabular}
  }
  \label{baseline}
\end{table}

From Table \ref{baseline}, we can observe the frequency content from the input is critical for low-resource setups. Comparing the SpecAugment configurations within the default setups, turning off frequency masking improves performance overall. This shows less significantly in the SWB 80hr setup (the model still improves on the in-domain set, but stagnates on the out-of-domain one). Comparing our MixSpeech setups, we observe the benefit of regularization on the time axis for the mixup. There is at least 7\% relative improvement on the evaluation sets across all datasets, which verifies our hypothesis on the benefits of regularization on the time axis for mixup-based methods (see Section 3.3). Finally, we turn off frequency masking in baselines and use \textit{Time enhanced} configuration for MixRep.

\begin{table}[t]
  \caption{WER on the WSJ corpus of proposed MixRep method. S denotes the set of layers to be selected from (see Section 3.3).}
  \label{t2}
  \centering
  \scalebox{0.9}{
  \begin{tabular}{ l c c | c c }
    \toprule
    \multirow{2}{*}{\centering Model} & \multicolumn{2}{c}{\textbf{With LM (\%)}} & \multicolumn{2}{c}{\textbf{No LM (\%)}}\\
     & dev93 & eval92 & dev93 & eval92 \\
    \midrule
    \textbf{Conformer} & & & & \\
    SpecAug. baseline & 6.2 & 4.3 & 10.4 & 7.7 \\
    + MixRep $S=\{0\}$ & 6.3 & 4.2 & 9.8 & 7.5 \\
    + MixRep $S=\{9\}$ & 6.1 & \textbf{4.1} & \textbf{9.4} & \textbf{7.2} \\
    + MixRep $S=\{0, 9\}$ & \textbf{6.0} & 4.2 & 9.8 & 7.5 \\
    \bottomrule
  \end{tabular}
  }
\end{table}

\subsection{Read English speech}

\begin{figure}[h]
    \centering
    \includegraphics[scale=0.11]{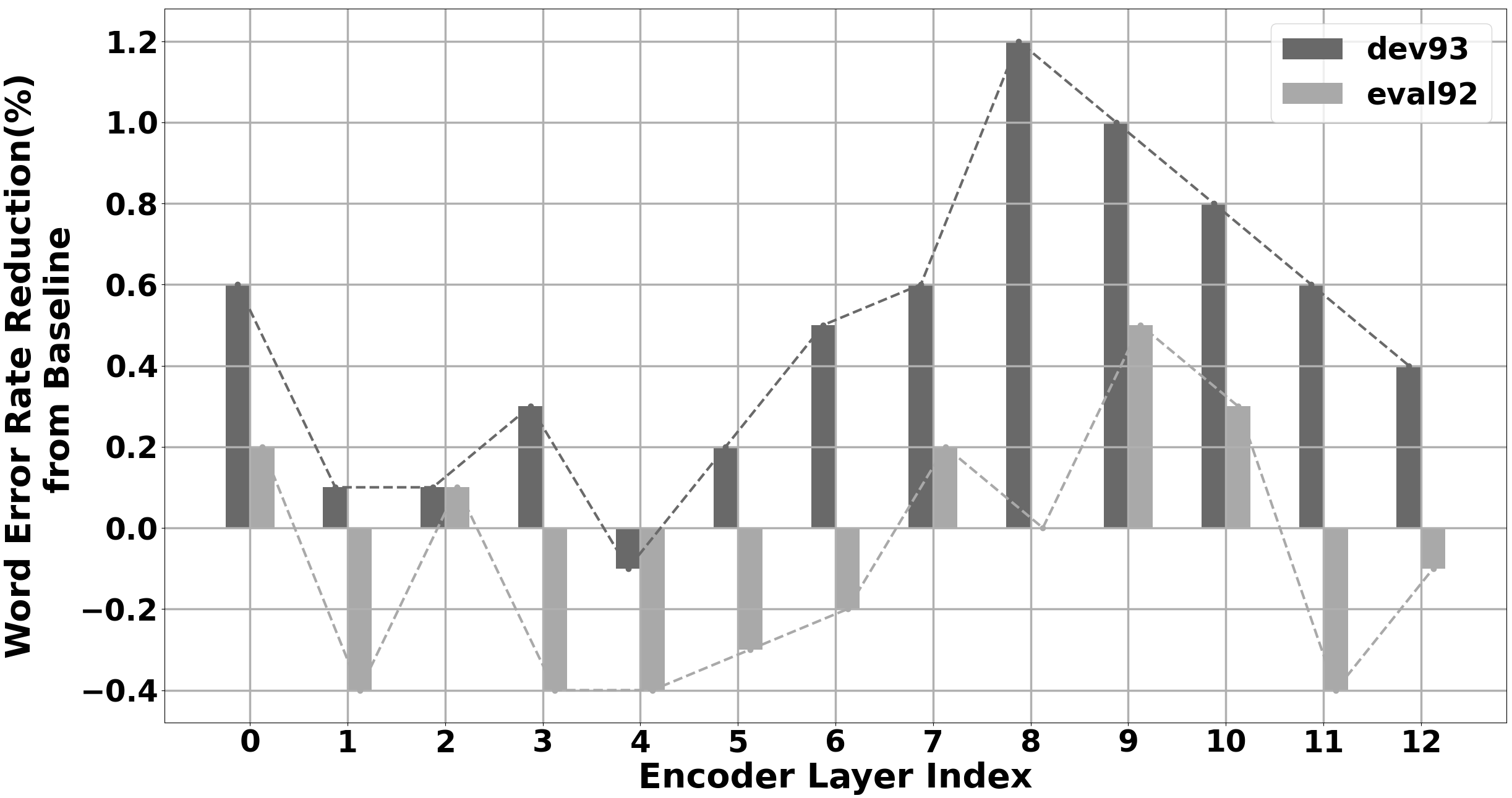}
    \caption{Per-layer improvement of MixRep compared to the SpecAugment baseline on the WSJ corpus.}
    \label{wsjlayer}
\end{figure}

 We compare MixRep to the best baseline and the input mixup for read English ASR. The results of MixRep applied at each layer are displayed in Figure \ref{wsjlayer}. The experimental results are illustrated in Table \ref{t2}.

 From Figure \ref{wsjlayer}, we observe mixing up in the deep layers (layer 7 to 10) gives good improvements over the baseline. This finding somewhat corresponds to the previous study \cite{belinkov2017analyzing}, which finds middle to deep layers of a CNN-RNN E2E ASR model trained on LibriSpeech contain more phonetic information than the early to middle layers. We hypothesize that certain layers of the E2E ASR model encode information similar to the output textual space, thus applying MixRep helps enforce this association by the linear relationship imposed.

We observe a superior performance using MixRep from the results presented in Table \ref{t3}. Mixing up the $9$-th layer representations outperforms the SpecAugment baseline by $+6.5\%$ relative and the input mixup by $+4\%$ on the evaluation set. When decoding with the LM, the improvement is diminished slightly, suggesting the benefits of the mixup may come from learning more linguistic knowledge in the encoder representations. 



\subsection{Spontaneous telephony speech}
\begin{figure}[h]
    \centering
    \includegraphics[scale=0.11]{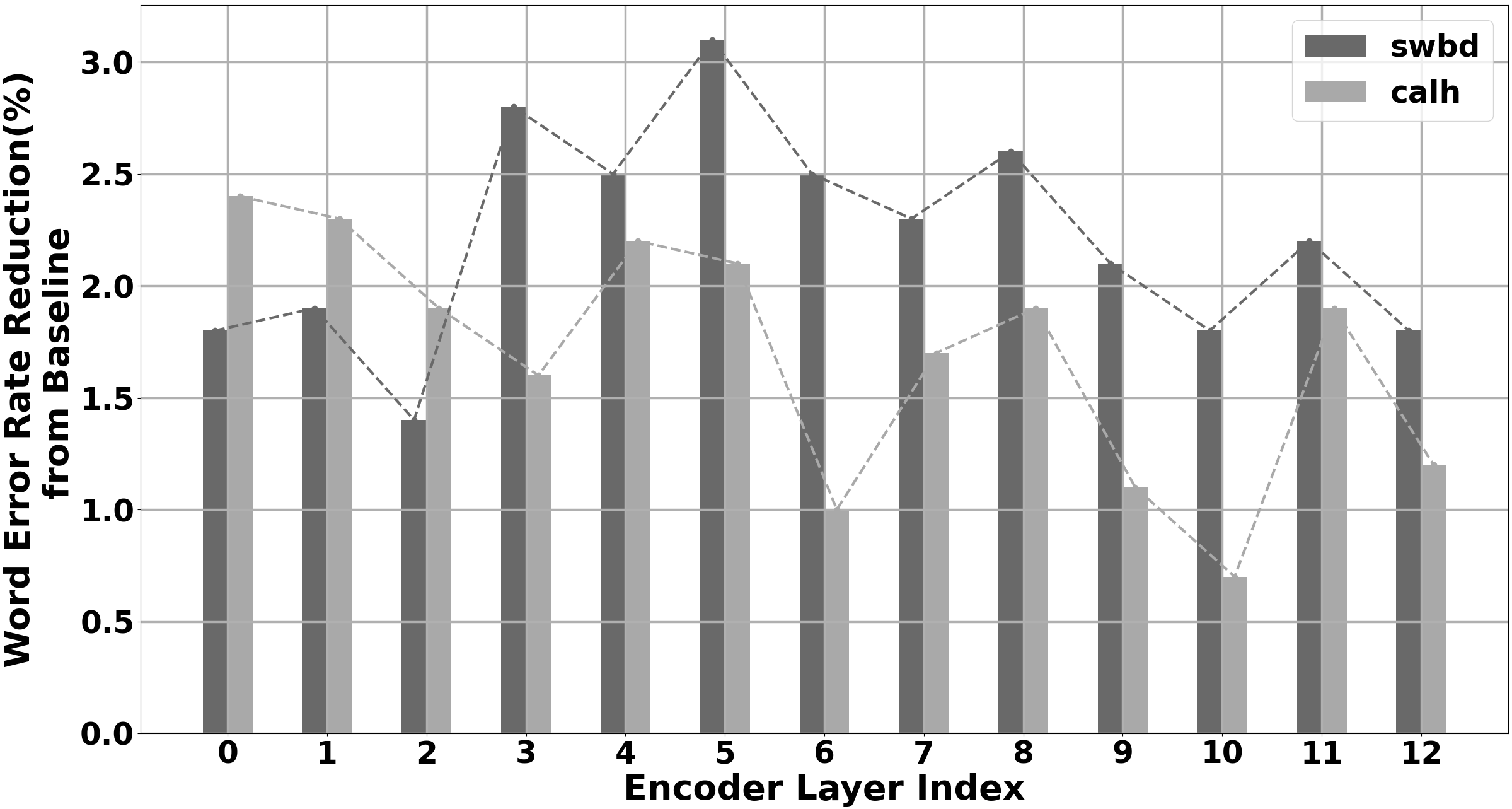}
    \caption{Per-layer improvement of MixRep compared to the SpecAugment baseline on the eval2000 using 40 hours of SWB.}
    \label{swblayer}
\end{figure}

We compare MixRep to other regularization methods for spontaneous telephony ASR. The results of MixRep applied at each layer are displayed in Figure \ref{swblayer}. The experimental results are illustrated in Table \ref{t3}.

From Figure \ref{swblayer}, we observe MixRep achieves significant and consistent gains over the SpecAugment baseline on the 40 hours SWB, which proves MixRep to be an effective method for low-resource training. Moreover, layer $5$, being the strongest performance on average, improves over the input mixup at the $0$-th layer. Compared to Figure \ref{wsjlayer}, we notice stronger improvements obtained by mixing up early to middle layer for the spontaneous telephony speech. Moreover, we spot a similar downward trend from layer 8 to layer 12, suggesting $\{8\}$ or $\{9\}$ can be a safe choice for the hyperparameter $S$.

For the SWB 40hr dataset in Table \ref{t3}, we verify applying MixRep to multiple layers can achieve better performance than a single layer. Mixing up both the $0$-th layer and $5$-th layer representations outperforms the SpecAugment baseline by a $+6.6\%$ relative on the Callhome set, suggesting complementary learning behavior upon regularizing multiple layers for ASR. This is similar to the previous finding for sound classification \cite{jindal20_interspeech}. For the SWB 80hr dataset in Table \ref{t3}, we observe the impact of training data size. The MixRep $S=\{0,5\}$ configuration leads the baseline by a +2.1\% relative after the training data is doubled. This verifies the data augmentation aspect of MixRep, but also shows the limitation of performance gain when the training data becomes sufficient. On the other hand, using the set $S=\{0,9\}$ outperforms $S=\{0,5\}$, which indicates the heuristic to select the optimal set $S$ is not optimal and is open for future work.

\begin{table}[t]
  \caption{WER on the eval'2000 using 40- and 80 hours training data subsets from the SWB corpus of proposed MixRep method.}
  \centering
  \label{t3}
  \scalebox{0.83}{
  \begin{tabular}{c l c c | c c }
    \toprule
    \multirow{2}{*}{\centering Train Data} & \multirow{2}{*}{\centering Model} & \multicolumn{2}{c}{\textbf{With LM (\%)}} & \multicolumn{2}{c}{\textbf{No LM (\%)}}\\
    & & swb & chm & swb & chm \\
   \midrule
    \multirow{5}{*}{\centering SWB 40hr} & \textbf{Conformer} & & & & \\
    & SpecAug. baseline  & 16.8 & 29.6 & 18.5 & 31.6 \\
    & + MixRep $S=\{0\}$ & 17.1 & 28.4 & 18.9 & 30.6 \\
    & + MixRep $S=\{5\}$ & \textbf{16.1} & 29.1 & \textbf{17.6} & 30.9 \\
    & + MixRep $S=\{0, 5\}$ & 16.3 & \textbf{27.7} & 17.7 & \textbf{29.5} \\
   \midrule
    \multirow{5}{*}{\centering SWB 80hr}
    & SpecAug. baseline  & 12.0 & 21.8 & 13.2 & 23.3 \\
    & + MixRep $S=\{0\}$ & 12.1 & \textbf{21.1} & 13.0 & 22.6 \\
    & + MixRep $S=\{0, 5\}$ & 11.9 & 21.3 & \textbf{12.8} & 22.8 \\
    & + MixRep $S=\{0, 9\}$ & \textbf{11.8} & 21.2 & \textbf{12.8} & \textbf{22.5} \\
    \bottomrule
  \end{tabular}
  }
\end{table}
\section{Conclusions}
In conclusion, we presented MixRep in this paper, a method to create artificial examples by interpolating hidden representations for E2E ASR training. We proposed an enhanced strategy for mixup-based methods, where a regularization along the time axis at the input is added. This is shown to be complementary to the feature regularization effect of the mixup for ASR. By experimenting on both read and spontaneous telephony styles of speech, we showed a significant and consistent improvement of MixRep over other regularization techniques such as SpecAugment and MixSpeech for low-resource ASR. We discussed the impact of training data size and the heuristic for searching the optimal set of eligible layers, which opens up future work.

\section{Acknowledgements}
The authors would like to thank Szu-Jui Chen for the meaningful discussion and suggestions on the work.
\bibliographystyle{IEEEtran}
\bibliography{mybib}

\end{document}